\documentclass[doublecol,figures,a4paper]{epl2} 

\usepackage{amsmath}
\pdfoutput=1

\title{Soft deformable self-propelled particles}

\author{A. M. Menzel\inst{1,2} \and T. Ohta\inst{2}}
\shortauthor{A. M. Menzel \etal}

\institute{
  \inst{1} {Institut f\"ur Theoretische Physik II: Weiche Materie, Heinrich-Heine-Universit\"at D\"usseldorf, 
    Universit\"atsstra{\ss}e 1, D-40225 D\"usseldorf, Germany}\\
  \inst{2} {Department of Physics, Graduate School of Science, Kyoto University, Kyoto 606-8502, Japan}
}


\abstract{
In this work we investigate the collective behavior of self-propelled particles that deform due to local pairwise interactions. We demonstrate that this deformation alone can induce alignment of the velocity vectors. The onset of collective motion is analyzed. Applying a Gaussian-core repulsion between the particles, we find a transition to disordered non-collective motion under compression. We here explain that this reflects the reentrant fluid behavior of the general Gaussian-core model now applied to a self-propelled system. Truncating the Gaussian potential can lead to cluster crystallization or more disordered cluster states. For intermediate values of the Gaussian-core potential we for the first time observe laning for deformable self-propelled particles. Finally, without the core potential, but including orientational noise, we connect our description to the Vicsek approach for self-propelled particles with nematic alignment interactions. 
}

\begin{document}

\maketitle

\section{Introduction}

The study of the onset of collective motion in systems of self-propelled particles was pioneered by Vicsek \textit{et al}.~\cite{vicsek1995novel}. In their model, point-like particles are assumed to move with constant magnitude of velocity in two spatial dimensions. The angular orientation of the velocity of each particle is subjected to stochastic noise. As a crucial further ingredient, polar alignment rules for the velocity orientations of neighboring particles are explicitly superimposed. When decreasing the noise amplitude or increasing the particle density, a macroscopic polar ordering transition occurs for the velocity orientations. 

This transition corresponds to the onset of collective motion. It was studied in more detail later \cite{gregoire2004onset,chate2008collective,bertin2009hydrodynamic}, also for the case of nematic alignment between the velocity vectors {\cite{peruani2008mean,ginelli2010large}} and for more complicated systems \cite{chate2008modeling,menzel2011collective}. 
Furthermore, systems of self-propelled particles that feature finite non-zero volume were investigated \cite{peruani2006nonequilibrium,baskaran2008enhanced}. In these studies it was demonstrated that steric interactions between rod-like particles can induce ordering. 

Here, we investigate the collective behavior in a system of self-propelled particles that can deform. For this purpose, we use an approach that was derived from symmetry arguments to model the behavior of a single self-propelled particle \cite{ohta2009deformable,tarama2011dynamics}. The description couples the velocity vector of the particle to a deformation tensor. This tensor characterizes the symmetric deformations starting from a spherical shape. Collective behavior within this model has been studied under explicitly imposed global coupling \cite{ohkuma2010deformable} or explicitly imposed pairwise alignment rules \cite{itino2011collective}. As an interesting feature, the breakdown of collective motion has been observed in the latter case under compression when an additional Gaussian repulsive potential was applied. 

In the following, we extend this model to the case in which pairwise interactions directly deform the particles. Imagine, for example, a scenario of soft self-propelled spheres that deform when squeezed together. Such objects will be called soft deformable self-propelled particles. We present a mechanism of velocity alignment mediated through such deformations. This can result in macroscopic ordered collective motion. It should be stressed that explicit alignment rules are not additionally imposed. 

We investigate the impact of this alignment mechanism on the onset of collective motion. Applying an anisotropic Gaussian-core potential between the particles, hexagonal translational order appears in the orientationally ordered collectively moving state. As was found for an isotropic Gaussian-core potential and an explicitly imposed alignment mechanism \cite{itino2011collective}, the state becomes disordered under compression. We now explain that this reflects the reentrant fluid behavior under compression generic for the Gaussian-core model. Truncating (and shifting) the Gaussian-core potential, we find variants of the disordered state under compression, such as an orientationally disordered cluster crystal and a cluster fluid-like state. 

Polar orientational velocity order in the collectively moving state results from a sufficient strength of the Gaussian-core potential. We observe laning behavior at intermediate values of the potential strength. Without the Gaussian-core potential, point-like particles as in the Vicsek model are considered. The implicit alignment mechanism then leads to nematic orientational order of the velocity vectors. Including orientational noise, we can link our description on soft deformable self-propelled particles to {corresponding Vicsek models \cite{peruani2008mean,ginelli2010large}}. Breakdown of orientational order is observed under increasing noise strengths.

\section{The model}

Here, we consider soft deformable self-propelled particles. That is, pairwisely repulsive interactions will directly lead to deformations of the particles. As mentioned above, a simple example is given by soft self-propelled spherical particles that deform when squeezed together. Because of the assumed particle softness, a soft repulsive potential force will be applied. 

Our model uses the same variables as introduced before \cite{ohta2009deformable}: a velocity vector $\mathbf{v}_i$ and a second-rank deformation tensor $\tens{S}_i$ for each particle $i$. $\tens{S}_i$ is symmetric and traceless. It can be related to the lowest-order symmetric deviation for each particle from a spherical to an elliptic shape \cite{ohta2009deformation}. 

The particle number is $N$ ($i=1\dots N$). Using these variables together with the position of each particle $\mathbf{r}_i$, we consider the following coupled equations of motion in two spatial dimensions: 
\begin{eqnarray}
\dot{\mathbf{r}}_i & = & \mathbf{v}_i, \label{req}\\[.1cm]
\dot{\mathbf{v}}_i & = & \gamma\,\mathbf{v}_i - \mathbf{v}_i^2 \mathbf{v}_i - a\,\tens{S}_i\cdot\mathbf{v}_i + \mu\,\mathbf{f}_i, \label{veq}\\
\dot{\tens{S}}_i & = & {}-\kappa\,\tens{S}_i 
           + b\,(1-2\,\tens{S}_i:\tens{S}_i)\Big(\mathbf{v}_i\mathbf{v}_i
               -\frac{1}{2}\mathbf{v}_i^2\tens{I}\Big) \nonumber\\
    & & {} - c\,(1-2\,\tens{S}_i:\tens{S}_i)\Big(\mathbf{f}_i\mathbf{f}_i
               -\frac{1}{2}\mathbf{f}_i^2\tens{I}\Big). \label{Seq}
\end{eqnarray}
Here, $\tens{I}$ is the unity matrix. We choose $\gamma>0$. {Then, in eq.~(\ref{veq}), the first two terms on the right enforce non-zero velocity values leading to self-propulsion in the stationary state. The first term on the right of eq.~(\ref{Seq})} induces relaxation back to the undeformed spherical state {for $\kappa>0$. Direct coupling between velocities and deformations is included through the terms with the coefficients $a$ and $b$.} In contrast to refs.~\cite{ohta2009deformable,itino2011collective}, we set $a<0$ and $b>0$. Then, {through these terms,} the velocity tends to orient parallel to the long axis of the elliptic deformation for each particle. Single interactionless particles move ballistically on straight trajectories. Unless stated otherwise, we set $\gamma=0.5$, $\kappa=1$, $a=-1$, and $b=0.5$. 

The force $\mathbf{f}_i$ acting on particle $i$ mimics soft steric interactions between the particles. It is included in the velocity equation with the coefficient $\mu$ as a soft core-to-core repulsive force. In contrast to ref.~\cite{itino2011collective}, we include anisotropy in the steric interaction force due to elliptic deformations. The soft interaction potential $U$ therefore contains the deformation tensor, 
\begin{equation}\label{U}
U = Ke^{{}-\frac{1}{2\sigma^2}\mathbf{r}_{ij}\cdot(\tens{I}-\tens{S}_i)\cdot(\tens{I}-\tens{S}_j)\cdot\mathbf{r}_{ij}},
\end{equation}
where $\mathbf{r}_{ij}=\mathbf{r}_{i}-\mathbf{r}_{j}$ 
and $\sigma>0$ is a characteristic interaction length. $\mathbf{f}_i$ follows as $\mathbf{f}_i=-\sum_{j\neq i}\partial U/\partial \mathbf{r}_{ij}$. The parameter $K$ is set to $1$ here since its effect is already included by $\mu$ (and $c$), and we choose $\sigma=1$. In realistic systems of self-propelled particles, the repulsive interaction potential is typically not long-ranged. We will therefore truncate the potential at a cut-off distance $r_c$ (and shift it). 

In the previous studies, the term with the coefficient $c$ in eq.~(\ref{Seq}) has not been considered. It is the key part of this work. We introduce this term to model soft deformable self-propelled particles. Then the particles directly deform due to soft repulsive steric interparticle interactions. For this purpose, we set $c>0$. At each time step, the deformation tensor for each particle can be parameterized as $\tens{S}_i=s_i(\mathbf{\hat{n}}_i\mathbf{\hat{n}}_i-\frac{1}{2}\tens{I})$. $s_i$ denotes the degree of deformation, the unit vector $\mathbf{\hat{n}}_i$ marks the long axis of deformation. $s_i=0$ corresponds to the undeformed spherical ground state and $s_i=2$ to a completely squeezed state. Here, we confine ourselves to medium deformations and limit the degree of deformation $s_i$ to the interval $s_i\in[0,1]$ through the prefactor $(1-2\,\tens{S}_i:\tens{S}_i)$. We stress again that there is no explicit alignment rule between the particle orientations as it was considered, e.g., in refs.~\cite{ohkuma2010deformable,itino2011collective}.

\section{Orientational alignment mechanism}\label{mechanism}

\begin{figure}
\onefigure[width=8.5cm]{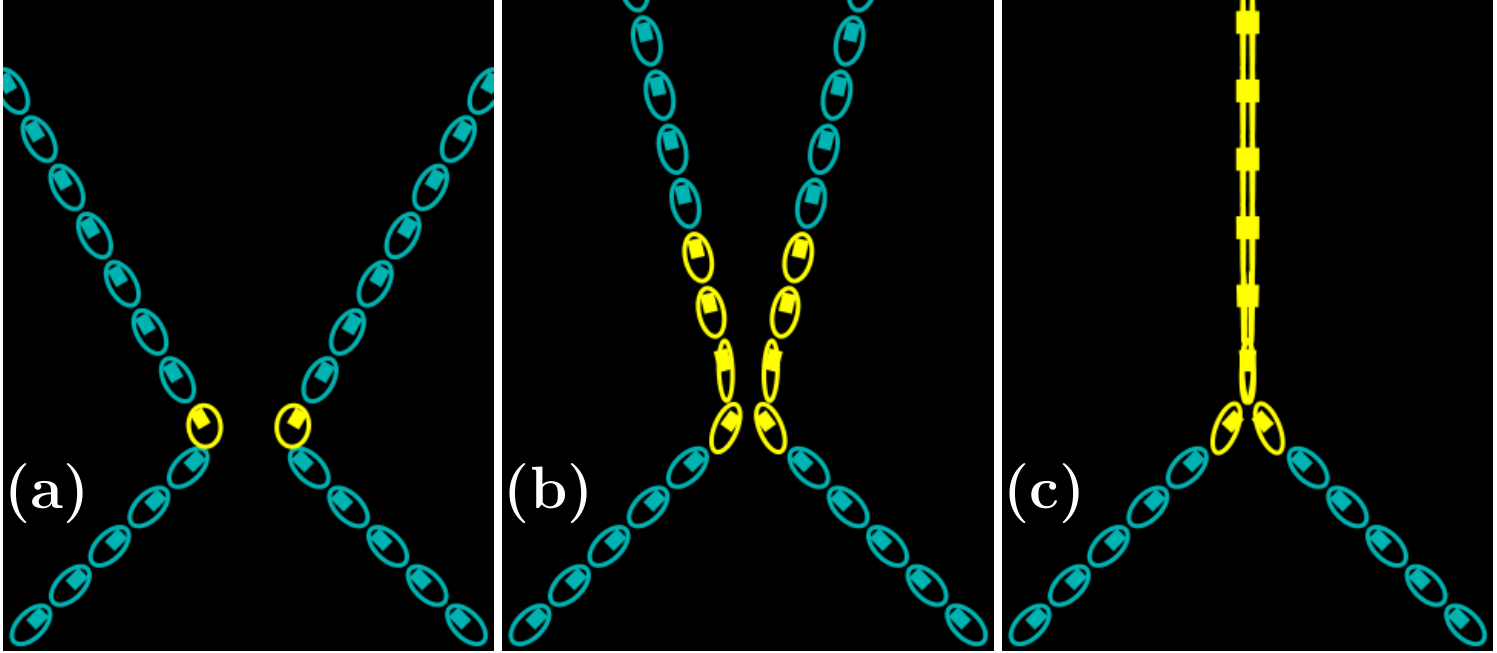}
\caption{Implicit alignment interaction for soft deformable self-propelled particles. Two particles are injected from the bottom and collide. A time series of particle states (bottom to top) is shown for each panel (a)--(c). The smaller the value of $\mu$ the more effective is the alignment interaction: (a) $\mu=1$, (b) $\mu=0.01$, and (c) $\mu=0$. States of particle separation smaller than $5\sigma$ are highlighted in brighter color.} 
\label{fig_mudep}
\end{figure}

First we explain the implicit alignment mechanism for soft self-propelled particles illustrated in fig.~\ref{fig_mudep}. 
The results were obtained by numerically integrating eqs.~(\ref{req})--(\ref{Seq}) forward in time for $c=20$. 
In the figure, deformation is indicated by the ellipsoidal shape of the particles. For this purpose, we parameterize the undeformed spherical boundary of particle $i$ by $\mathbf{R_0}_i(\varphi)=R_0(\cos(\varphi),\sin(\varphi))$ with $R_0$ the radius and $\varphi\in[0,2\pi[$. The boundary location of the deformed state is drawn via $\mathbf{R}_i(\varphi)=(\tens{I}+{\alpha}\tens{S}_i)\cdot\mathbf{R_0}_i(\varphi)$ with an extra factor of ${\alpha}=2$ used for illustration only in the figures. 
Orientation and magnitude of the velocity vectors are marked by the orientation and length of the short thick lines within the ellipsoids, respectively. To draw the figures, we adjust the ground state radius $R_0$ of the undeformed particles for best visualization. 

When two particles come close to each other (``collide''), they deform due to steric interactions modeled by the term with the coefficient $c$ in eq.~(\ref{Seq}). For $c>0$ the particles are compressed along the connecting line between their centers of mass. Consequently the particles elongate perpendicularly to that direction. Since $a<0$ and $b>0$, the velocity vector and long axis of deformation tend to align parallel to each other for each particle. Together, these two mechanisms lead to an effective alignment of the velocity vectors of the two particles. The smaller the repulsive force in eq.~(\ref{veq}), the more efficient is the alignment mechanism as illustrated in figs.~\ref{fig_mudep} (a)--(c).

\section{Polar orientational alignment}

\begin{figure}
\onefigure[width=8.5cm]{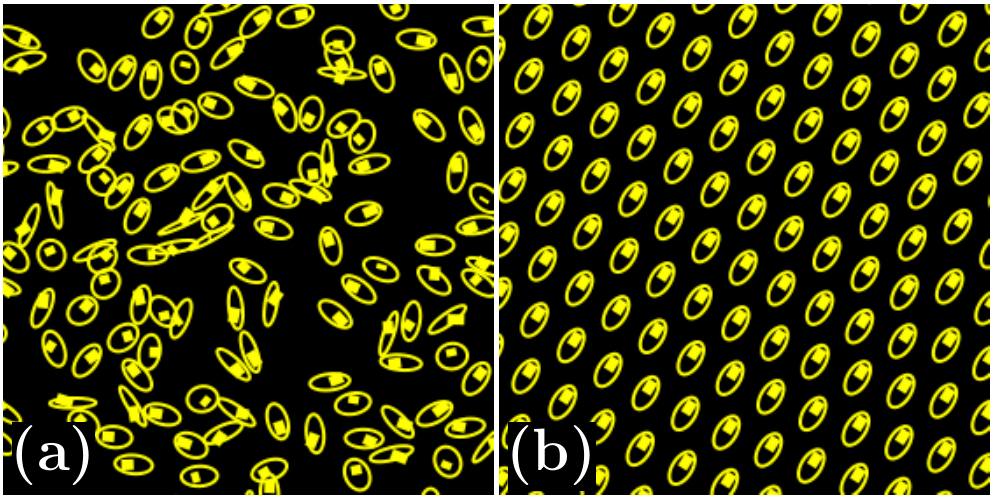}
\caption{Snapshots from the ordering process of a representative sample, starting from a disordered state of random initial conditions. The sample is still in the disordered state in (a) at $t=600$, but has orientationally and translationally ordered in (b) at $t=12500$. System parameters: $N=300$, $\rho=0.14$, $\mu=0.01$, $c=2.5$.} 
\label{fig_orderedstate}
\end{figure}

To investigate the collective behavior of soft deformable self-propelled particles, we first study the ordering process in an initially disordered crowd. For this purpose, $N$ particles were simulated in a periodic box of size $L\times L$. In the numerical calculations, to obtain box lengths of order unity, all lengths were rescaled by a factor of $100\sigma$. A measure for the density is defined by $\rho=N\sigma^2/L^2$. As for all subsequently reported results, we started from a random spatial distribution, random values for the velocity orientations and magnitudes, and vanishing deformation. Snapshots from the ordering process for a representative sample can be found in fig.~\ref{fig_orderedstate} for the given material parameters (only part of the numerical calculation box is shown). 

It is important to note that the alignment mechanism in eq.~(\ref{Seq}) through the force term with the coefficient $c$ alone would lead to nematic orientational alignment. From fig.~\ref{fig_orderedstate} (b), however, we find polar orientational alignment of the velocity vectors. This polar alignment must therefore result from the core repulsion in eq.~(\ref{veq}) with coefficient $\mu$. Also the hexagonal translational order must be attributed to the core repulsion. Both conclusions will be verified. 

To analyze the role of the implicit alignment mechanism, the time it takes until the system orientationally orders was measured as a function of $c$. For this purpose, $N=140$ particles were simulated in a periodic box with density $\rho=0.24$. Other system parameters were the same as used to obtain fig.~\ref{fig_mudep}, except for $\mu=0.05$. 

\begin{figure}
\onefigure[width=8.cm]{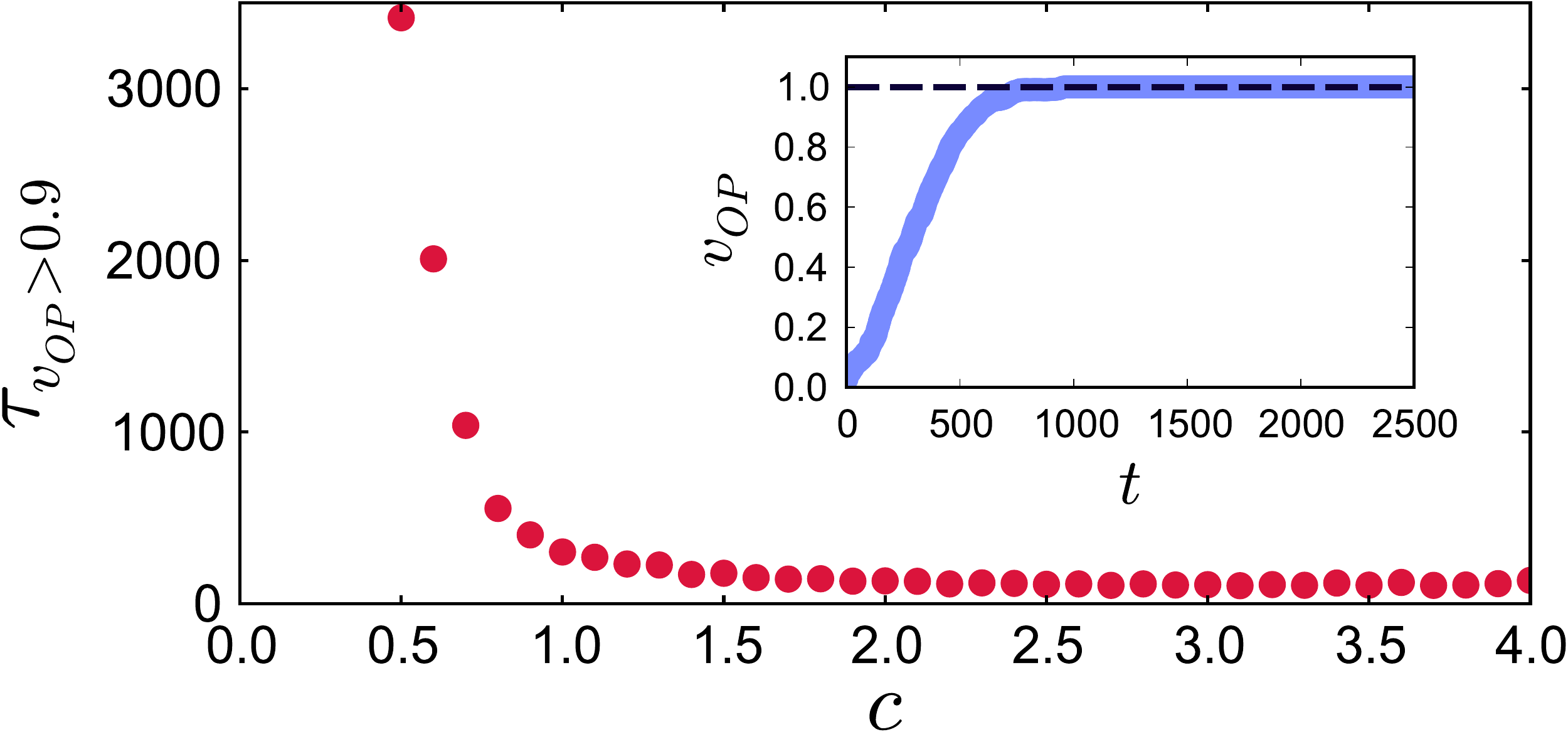}
\caption{Time $\tau_{v_{OP}>0.9}$ until a polar orientational degree of velocity order $v_{OP}$ of $0.9$ is reached as a function of the deformational force strength $c$. Each data point is an average over $100$ samples starting from a different disordered state of random initial conditions. The inset shows a typical time evolution of the sample averaged $v_{OP}$ (here for $c=0.8$). System parameters: $N=140$, $\rho=0.24$, $\mu=0.05$.} 
\label{fig_timedepordering}
\end{figure}

As an order parameter, we evaluated the magnitude of the sample-averaged velocity orientation, $v_{OP}=\|\frac{1}{N}\sum_{i=1}^N\mathbf{v}_i/v_i\|$. We consider the sample as being ordered when $v_{OP}>0.9$. Fig.~\ref{fig_timedepordering} illustrates as a function of $c$ the time $\tau_{v_{OP}>0.9}$ that it takes for the sample to orientationally order. Each data point is an average over $100$ samples of different initial conditions. It is obvious that alignment through deformation strongly supports the onset of collective motion. The ordering time $\tau_{v_{OP}>0.9}$ seems to diverge for small values of $c$ for the given parameter values. 
In the inset of the figure we show the ordering process as a function of time for one specific value of $c$. 

\begin{figure}
\onefigure[width=8.cm]{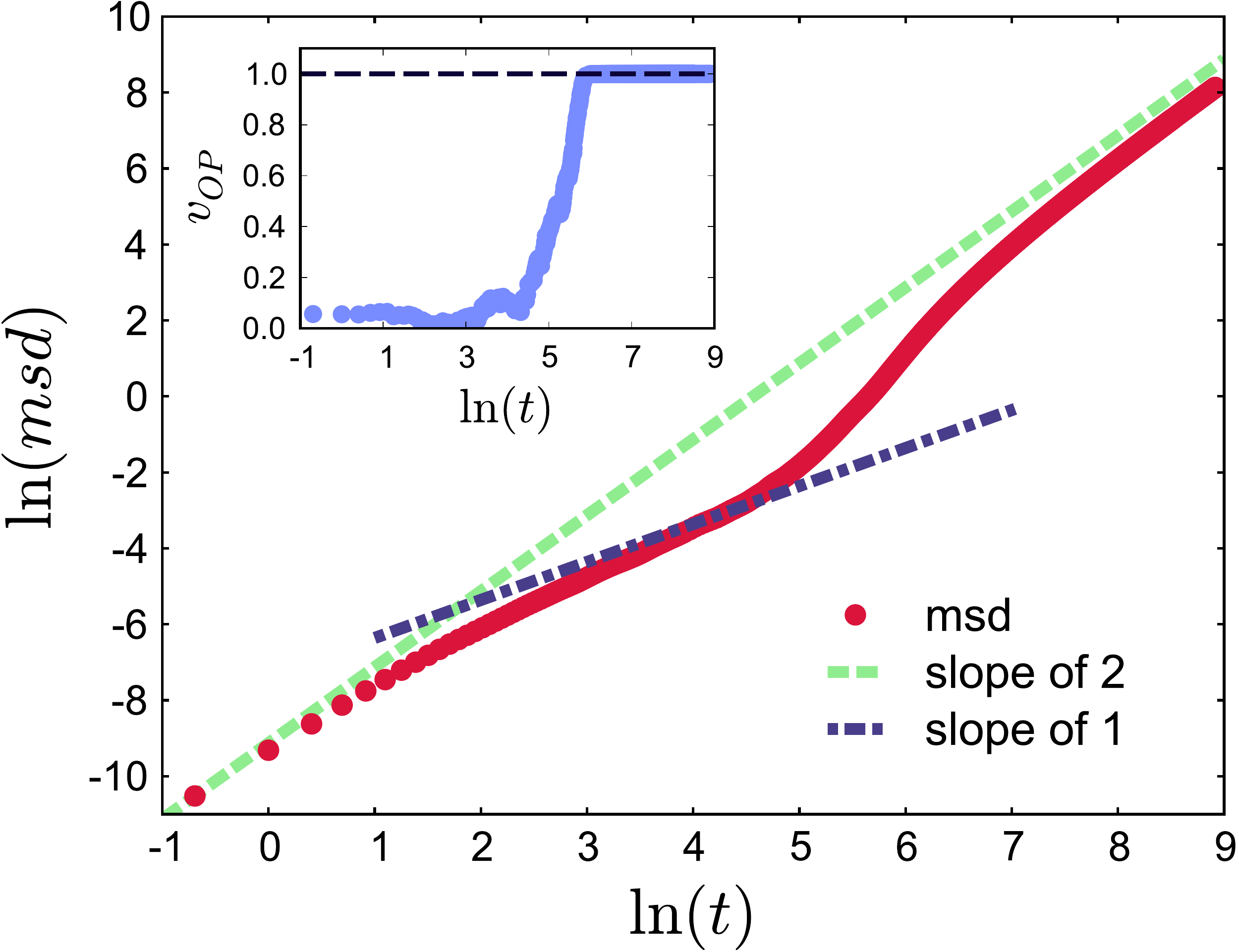}
\caption{Time evolution of the mean square displacement (msd) plotted on double logarithmic scales reflects the ordering process. The evolution started from random initial conditions. In the disordered state the slope of the msd data curve decreases (a line of slope $1$ is included for comparison). Long-ranged ordering starts around the point where the curvature becomes positive. Asymptotically, the slope approaches $2$ in the ordered collectively moving state. The inset shows the degree of polar orientational velocity order $v_{OP}$ for comparison. System parameters as given in the caption of fig.~\ref{fig_orderedstate}.} 
\label{fig_msdonset}
\end{figure}

The ordering process can also be tracked by the time evolution of the mean square displacement. An example is given by fig.~\ref{fig_msdonset} in a log-log plot. 
Initially, before the first collisions occur, we find a slope close to $2$ indicating ballistic motion. There follows a time interval of slope closer to $1$ when particle velocities are not yet orientationally ordered. Therefore collisions are frequent, which leads to an overall behavior close to diffusion. We observe a kink in the curve when collective motion sets in. Finally, all particles move into the same direction and collisions are rare. Consequently the motion of each particle is close to ballistic again. As a result, the slope of the mean square displacement asymptotically approaches a value of $2$. 

In a previous study on deformable self-propelled particles, it was found that under compression the hexagonal translational and polar orientational order breaks down when a critical density is reached \cite{itino2011collective}. The order reappeared under subsequent expansion. A hysteretic effect was numerically observed. As an interparticle potential the isotropic analog of eq.~(\ref{U}) was used. 

We investigated this effect in detail and found that deformability of the particles is not necessary to observe it. If we fix the particle shape, we can still observe the breakdown of order under compression. This is the case even for spherical particles. However, increasing the deformational interactions via higher values of $c$ supports the effect, and the breakdown of order occurs at lower densities. 

As a consequence, we identify the soft interaction potential of Gaussian form as the main source of the reported effect. In the equilibrium case, Gaussian-core systems have been studied in detail \cite{stillinger1976phase,prestipino2005phase,prestipino2011hexatic}, also for anisotropic forms of the Gaussian interaction potential \cite{prestipino2007phase} and anisotropic particles \cite{rex2007dynamical}. At high density, the corresponding phase diagram features melting of crystalline phases into a fluid. Since the system is typically fluid at low densities as well, this high-density fluidity is called reentrant. We therefore conclude that we observe the analog of this reentrant fluid behavior in self-propelled Gaussian-core systems. 

\begin{figure}
\onefigure[width=8.5cm]{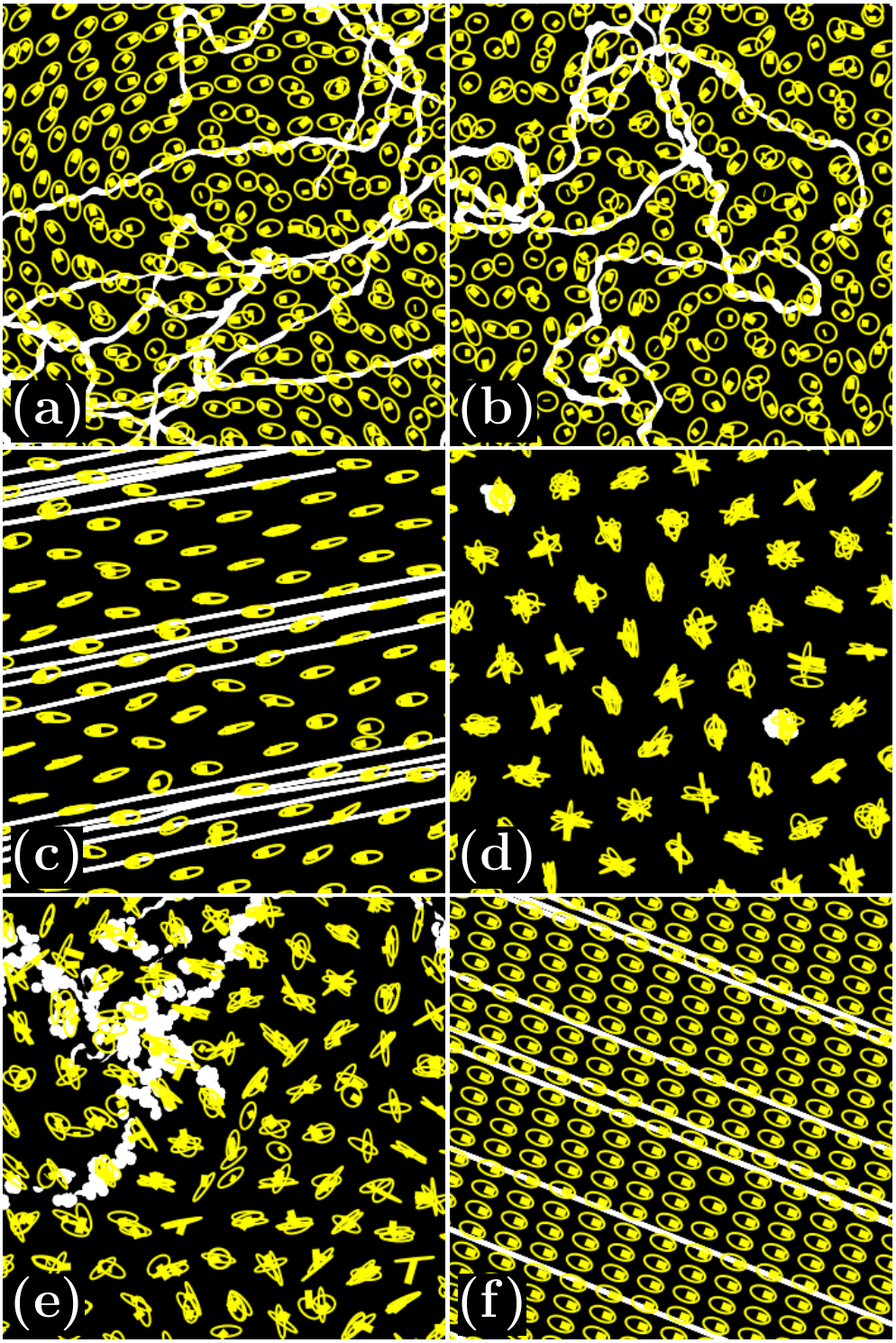}
\caption{Illustration of the collective behavior at high densities after sufficiently slow compression from a hexagonally ordered polar state. The initial density was $\rho=0.2$ in all cases. Trajectories of selected particles are indicated in white (line thickness decreases with increasing velocity). Different panels correspond to different cut-off radii $r_c$ and densities: (a) $r_c=5\sigma$, $\rho=0.84$, (b) $r_c=5\sigma$, $\rho=0.93$, (c) $r_c=2\sigma$, $\rho=0.76$, (d) $r_c=2\sigma$, $\rho=1.27$, (e) $r_c=1.5\sigma$, $\rho=1.27$, (f) $r_c=2.5\sigma$, $\rho=0.37$. System parameters: $N=1000$, $\mu=0.05$, $c=2.5$.} 
\label{fig_breakdown}
\end{figure}

In fig.~\ref{fig_breakdown} we qualitatively show examples of the behavior that we found at high densities. 
We varied the cut-off distances $r_c$ of the Gaussian-core potential. 
At the initial density $\rho=0.2$, all samples showed hexagonal polar order analogous to fig.~\ref{fig_orderedstate} (b). Example trajectories of single particles are indicated in white. The local line thickness decreases with increasing magnitude of the recorded velocity. Only part of the numerical calculation box is shown. 

Figs.~\ref{fig_breakdown} (a) and (b) correspond to a cut-off distance $r_c=5\sigma$. After global order has broken down, we still find some local order (a), which vanishes under further compression (b). The velocity magnitudes along a trajectory in (b) are highly non-homogeneous, indicating frequent collisions. This scenario is identified with the reentrant fluid behavior outlined above for the Gaussian-core model. 

Interesting behavior occurs at smaller cut-off lengths. Figs.~\ref{fig_breakdown} (c) and (d) show an example for $r_c=2\sigma$. Again the softness of the potential plays an important role. It allows putting several particles at the same spatial location with only finite energetic penalty. Under slow compression, this happens already when the system is still in the ordered crystalline state (c). Several lattice points are occupied by more than one particle. The motion is still orientationally ordered as indicated by the straight trajectories. At higher densities this order breaks down. We obtain a cluster crystal \cite{likos1998freezing,schmidt1999an}, here in a system of self-propulsion, with several particles stacked at each lattice site (d). The deformed particles are not ordered orientationally. As indicated by the localized trajectories, the particles are trapped at the lattice points and the clusters are immobile. The situation may be connected to the effect that particles in a plane under compression stack on top of each other at high densities. Due to this effect, the area density $\rho$ can become larger than $1$. 

At still smaller cut-off lengths $r_c=1.5\sigma$, this structure is more mobile, as indicated by the trajectories in fig.~\ref{fig_breakdown} (e). In effect, the particles can move over the plane. We may call this scenario a ``cluster fluid-like'' state. 

A special situation was found for $r_c=2.5\sigma$. Before melting, the hexagonal crystal structure turned into a rectangular one. The latter is still orientationally ordered, as can be seen from the straight trajectories in fig.~\ref{fig_breakdown} (f). 

Slowly expanding the systems to their initial density, all of these structures turned back into the polar hexagonal crystal structure. We also observed hysteretic effects \cite{itino2011collective}.

\section{Laning}

Laning has not been observed previously for deformable self-propelled particles. 
{For rigid self-propelled particles, it has been reported very recently \cite{mccandlish2012spontaneous,wensink2012emergent}.} 
The laning state is characterized by globally nematic, but locally polar alignment of the velocity vectors. We have already seen that high values of the strength $\mu$ of the Gaussian-core repulsion lead to global polar order. In the next section, we will see that without the Gaussian-core repulsion, $\mu=0$, nematic order is obtained globally and locally. Thus the laning state should be found for intermediate values of the core repulsion strength $\mu$. In addition, the system parameters need to be chosen to give a high aspect ratio of the typical particle deformation. 

\begin{figure}
\onefigure[width=8.5cm]{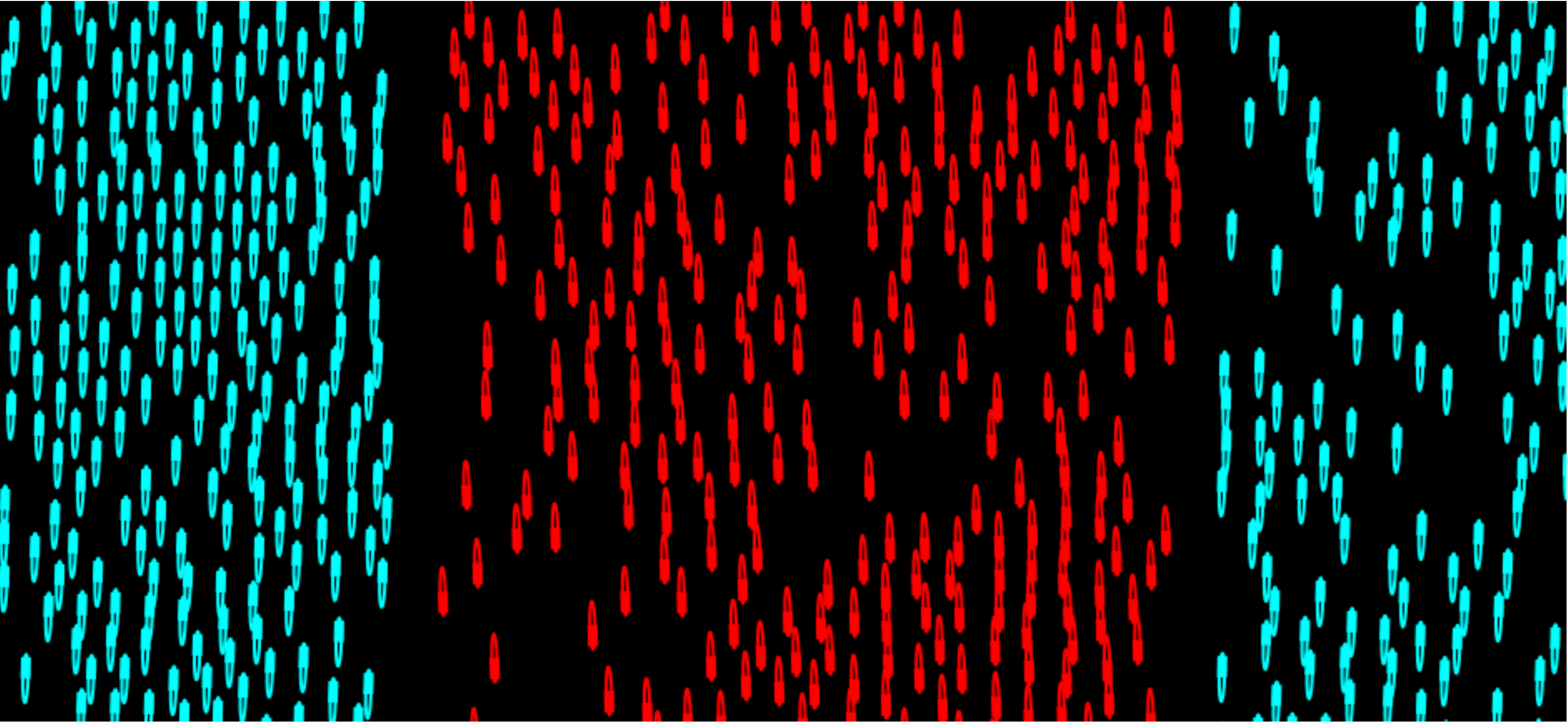}
\caption{Laning in a system of soft deformable self-propelled particles of high aspect ratio. Red particles move downward, turquoise ones upward. Only part of the numerical calculation box is shown. System parameters: $N=1200$, $\rho=0.14$, $a = -10$, $\mu=0.03$, $c = 1.2$.} 
\label{fig_laning}
\end{figure}

At this stage, we only point out that laning can indeed be observed under these circumstances. Fig.~\ref{fig_laning} shows an example. 
Again we started from random initial conditions to obtain this result. The velocity alignment is nematic on a global scale and polar within each lane. 
Nonzero but not too high values of $c>0$ seem to support the macroscopic separation into different lanes. In the example of fig.~\ref{fig_laning}, we measured a sample-averaged degree of deformation $\langle s \rangle\approx0.8$. This corresponds to an aspect ratio of about $2.33:1$.

\section{Nematic orientational alignment}

\begin{figure}
\onefigure[width=8.5cm]{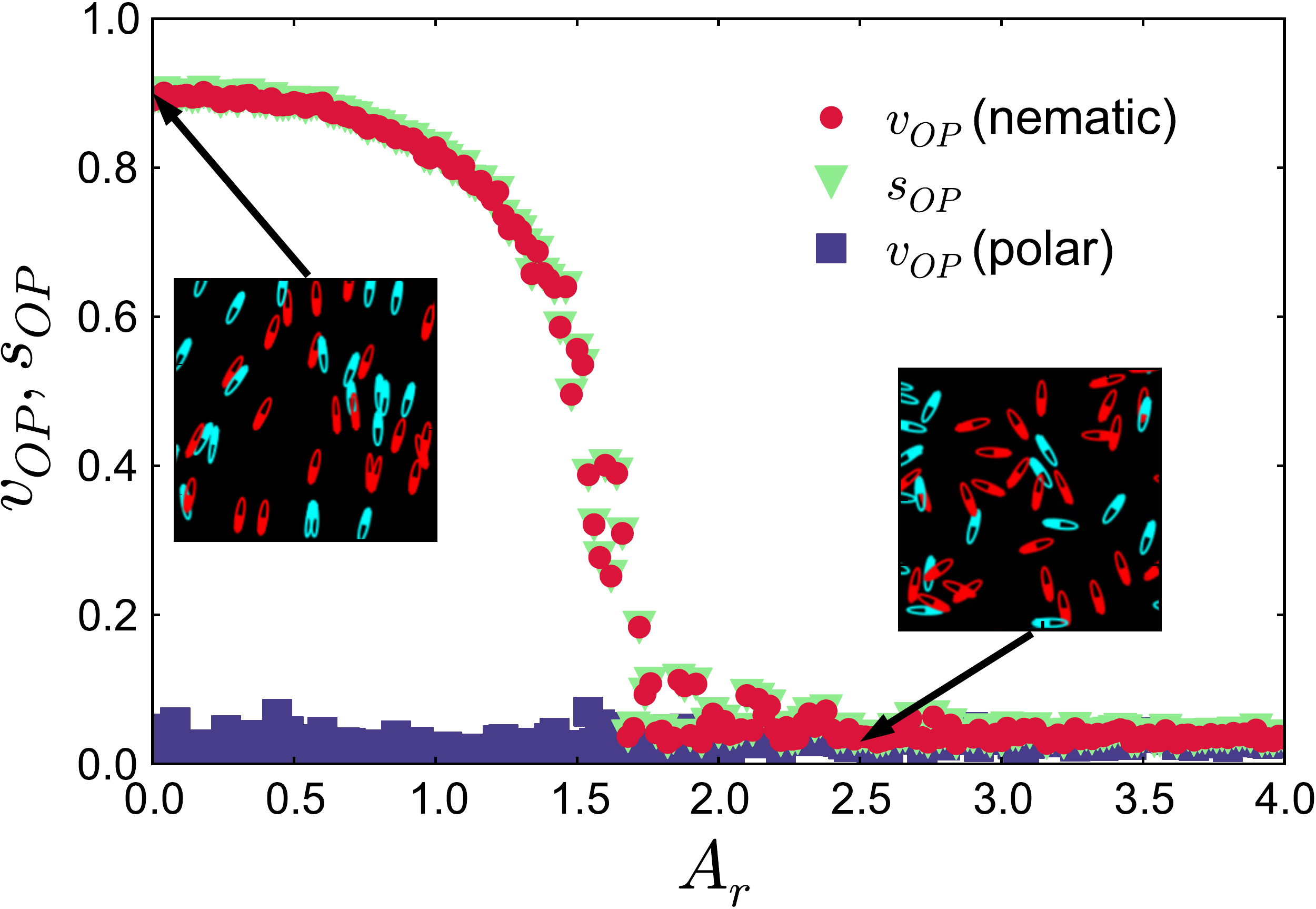}
\caption{Nematic and polar orientational velocity order parameters $v_{OP}$ and orientational deformational order parameter $s_{OP}$ as a function of orientational deformational noise strength $A_r$. Left and right insets illustrate nematic alignment below and orientational disorder above the order-disorder transition, respectively. Red particles move downward, turquoise ones upward; only a very small portion of the sample is depicted. System parameters: $N=800$, $\rho=0.09$, $a = -5$, $\mu=0$, $c = 0.5$.} 
\label{fig_vicsek}
\end{figure}

Finally, our scope is to connect our description of soft deformable self-propelled particles with implicit alignment mechanism to the Vicsek model. As a first step, we set the steric core-to-core repulsion to zero ($\mu=0$). We thus consider sterically point-like particles. The only remaining alignment mechanism via the term with the coefficient $c$ is of nematic character. It should lead to nematic orientational order of the particle velocities. Indeed we observe such a scenario as shown, for example, by the left inset in fig.~\ref{fig_vicsek}. 

Second, we have to introduce orientational noise into our equations. It turns out that noise in the orientation of the deformation tensors $\tens{S}_i$ is much more effective than noise in the orientation of the velocity vectors $\mathbf{v}_i$. 
Therefore, at each time step and for each particle $i$ we pick a random angle $\xi_i$ from a Gaussian distribution and multiply it by a constant factor $A_r$. The deformation tensor $\tens{S}_i$ is then rotated by the angle $A_r\xi_i$. In this way we introduce purely orientational noise, without changing the degree of deformation. $A_r$ sets the noise strength.

An example plot of the orientational order parameters as a function of the noise strength $A_r$ is shown in fig.~\ref{fig_vicsek}. 
As expected from the {Vicsek model with nematic alignment interactions \cite{peruani2008mean, ginelli2010large}}, the nematic orientational velocity order parameter goes to zero at a critical noise strength. At least for the finite system sizes investigated, the transition appears to be continuous. {However, it is not possible to draw a final conclusion about the nature of the transition because a discontinuity may emerge at very high particle numbers \cite{chate2008collective,ginelli2010large}.} At all noise strengths, the degree of deformational order is practically of the same magnitude as the degree of nematic velocity order. As expected, the polar velocity order is close to zero for all values of $A_r$. We have thus linked the scenario of soft deformable self-propelled particles to the {corresponding} Vicsek approach.

\section{Conclusions}

We introduced a description of soft deformable self-propelled particles. For this purpose, we included into the equations of motion a pairwise repulsive force that can directly deform the particle shape. As demonstrated this leads to an implicit pairwise velocity alignment mechanism. The latter can significantly accelerate the process of long-ranged orientational ordering. 

Furthermore, we studied the breakdown of collective motion under compression for the case of the introduced anisotropic repulsive Gaussian-core potential. This transition is identified as the analog to the reentrant fluid behavior in Gaussian-core equilibrium systems \cite{stillinger1976phase,prestipino2005phase,prestipino2011hexatic,prestipino2007phase}. The new soft deformational interactions were observed to shift this transition to lower densities. Interesting scenarios such as rectangular translational order, cluster crystals, and less ordered cluster states were observed at small cut-off lengths for the Gaussian interaction potential. We are not aware of any previous studies on cluster crystals or crystals of rectangular lattice structure for self-propelled particles. 
In addition, we for the first time found laning in deformable self-propelled particle systems. 

{When strong Gaussian-core repulsion enforces translational crystalline order, only polar orientational velocity order is compatible with the imposed lattice structure. At very low values of the core repulsion, nematic velocity order emerged from our implicit alignment mechanism. Laning was found for intermediate values as a state of global nematic but local polar orientational velocity order.}

Finally, we linked our characterization to the Vicsek approach for point-like particles with nematic alignment interactions. 
Increasing the noise of deformation orientations caused the nematic orientational order to break down at a critical noise strength. 

Our approach should be useful for the investigation of soft biological tissue. As an example we mention soft motile cells that migrate on flat substrates \cite{rappel1999self}. Possible additional adhesive forces can be included into our description. Another example may be given by soft vesicles that deform but do not merge when approaching each other. Candidates for non-biological model systems are self-propelled droplets at gas-liquid interfaces \cite{chen2009self} or in bulk fluid \cite{thutupalli2011swarming}. These droplets were demonstrated to deform under steric confinement interactions \cite{chen2009self} and to survive collisions \cite{thutupalli2011swarming}. 
Finally, we wish to point out the qualitative similarity in appearance between our alignment mechanism illustrated in fig.~\ref{fig_mudep} and the process of inelastic collisions in general. The latter was recently outlined as a bridge connecting particle-based descriptions of the Vicsek-type to the characterization of granular media \cite{grossman2008emergence}.

\acknowledgments{
The authors thank Hartmut L\"owen for a stimulating discussion. Support from the Deutsche Forschungsgemeinschaft through project ``Nichtgleichgewichtsph\"anomene in Weicher Materie/Soft Matter'' no.~LO 418/15-1 and from the JSPS Core-to-Core Program ``International research network for non-equilibrium dynamics of soft matter'' is gratefully acknowledged. 
}


\begin{thebibliography}{10}
\expandafter\ifx\csname url\endcsname\relax\def\url#1{\texttt{#1}}\fi

\bibitem{vicsek1995novel}
\Name{Vicsek T., Czir{\'o}k A., Ben-Jacob E., Cohen I. \and Shochet O.}
  \REVIEW{Phys. Rev. Lett.}{75}{1995}{1226}.

\bibitem{gregoire2004onset}
\Name{Gr{\'e}goire G. \and Chat{\'e} H.} \REVIEW{Phys. Rev.
  Lett.}{92}{2004}{025702}.

\bibitem{chate2008collective}
\Name{Chat{\'e} H., Ginelli F., Gr{\'e}goire G. \and Raynaud F.} \REVIEW{Phys.
  Rev. E}{77}{2008}{046113}.

\bibitem{bertin2009hydrodynamic}
\Name{Bertin E., Droz M. \and Gr{\'e}goire G.} \REVIEW{J. Phys. A: Math.
  Theor.}{42}{2009}{445001}.

\bibitem{peruani2008mean}
\Name{Peruani F., Deutsch A. \and B{\"a}r M.} \REVIEW{Eur. Phys. J. Special
  Topics}{157}{2008}{111}.

\bibitem{ginelli2010large}
\Name{Ginelli F., Peruani F., B{\"a}r M. \and Chat{\'e} H.} \REVIEW{Phys. Rev.
  Lett.}{104}{2010}{184502}.

\bibitem{chate2008modeling}
\Name{Chat{\'e} H., Ginelli F., Gr{\'e}goire G., Peruani F. \and Raynaud F.}
  \REVIEW{Eur. Phys. J. B}{64}{2008}{451}.

\bibitem{menzel2011collective}
\Name{Menzel A.~M.} \REVIEW{Phys. Rev. E}{85}{2012}{021912}.

\bibitem{peruani2006nonequilibrium}
\Name{Peruani F., Deutsch A. \and B{\"a}r M.} \REVIEW{Phys. Rev.
  E}{74}{2006}{030904(R)}.

\bibitem{baskaran2008enhanced}
\Name{Baskaran A. \and Marchetti M.~C.} \REVIEW{Phys. Rev.
  Lett.}{101}{2008}{268101}.

\bibitem{ohta2009deformable}
\Name{Ohta T. \and Ohkuma T.} \REVIEW{Phys. Rev. Lett.}{102}{2009}{154101}.

\bibitem{tarama2011dynamics}
\Name{Tarama M. \and Ohta T.} \REVIEW{Eur. Phys. J. B}{83}{2011}{391}.

\bibitem{ohkuma2010deformable}
\Name{Ohkuma T. \and Ohta T.} \REVIEW{Chaos}{20}{2010}{023101}.

\bibitem{itino2011collective}
\Name{Itino Y., Ohkuma T. \and Ohta T.} \REVIEW{J. Phys. Soc.
  Jpn.}{80}{2011}{033001}.

\bibitem{ohta2009deformation}
\Name{Ohta T., Ohkuma T. \and Shitara K.} \REVIEW{Phys. Rev.
  E}{80}{2009}{056203}.

\bibitem{stillinger1976phase}
\Name{Stillinger F.~H.} \REVIEW{J. Chem. Phys.}{65}{1976}{3968}.

\bibitem{prestipino2005phase}
\Name{Prestipino S., Saija F. \and Giaquinta P.~V.} \REVIEW{Phys. Rev.
  E.}{71}{2005}{050102}.

\bibitem{prestipino2011hexatic}
\Name{Prestipino S., Saija F. \and Giaquinta P.~V.} \REVIEW{Phys. Rev.
  Lett.}{106}{2011}{235701}.

\bibitem{prestipino2007phase}
\Name{Prestipino S. \and Saija F.} \REVIEW{J. Chem. Phys.}{126}{2007}{194902}.

\bibitem{rex2007dynamical}
\Name{Rex M., Wensink H.~H. \and L\"owen H.} \REVIEW{Phys. Rev.
  E.}{76}{2007}{021403}.

\bibitem{likos1998freezing}
\Name{Likos C.~N., Watzlawek M. \and L{\"o}wen H.} \REVIEW{Phys. Rev.
  E}{58}{1998}{3135}.

\bibitem{schmidt1999an}
\Name{Schmidt M.} \REVIEW{J. Phys.: Condens. Matter}{11}{1999}{10163}.

\bibitem{mccandlish2012spontaneous}
\Name{McCandlish S.~R., Baskaran A. \and Hagan M.~F.} \REVIEW{Soft
  Matter}{8}{2012}{2527}.

\bibitem{wensink2012emergent}
\Name{Wensink H.~H. \and L{\"o}wen H.} \REVIEW{Arxiv preprint
  arXiv:1204.0381}{}{2012}{}.

\bibitem{rappel1999self}
\Name{Rappel W.~J., Nicol A., Sarkissian A., Levine H. \and Loomis W.~F.}
  \REVIEW{Phys. Rev. Lett.}{83}{1999}{1247}.

\bibitem{chen2009self}
\Name{Chen Y.-J., Nagamine Y. \and Yoshikawa K.} \REVIEW{Phys. Rev.
  E}{80}{2009}{016303}.

\bibitem{thutupalli2011swarming}
\Name{Thutupalli S., Seemann R. \and Herminghaus S.} \REVIEW{New J.
  Phys.}{13}{2011}{073021}.

\bibitem{grossman2008emergence}
\Name{Grossman D., Aranson I.~S. \and Ben~Jacob E.} \REVIEW{New J.
  Phys.}{10}{2008}{023036}.

\end{thebibliography}

\end{document}